\documentclass[a4paper,11pt]{article}

\usepackage{jheppub} 
\usepackage[T1]{fontenc} 

\title{Slow-Goldstone mode generated by order from quantum disorder
and its experimental detection}

\author[a,b]{Fadi Sun}
\author[a,b,c]{and Jinwu Ye}
\affiliation[a]{Tsung-Dao Lee Institute, Shanghai 200240, China }
\affiliation[b]{Institute for Quantum Science and Engineering, Shenzhen 518055, China }
\affiliation[c]{Department of Physics and Astronomy,
Mississippi State University, Mississippi State, MS 39762, USA}

\abstract{
The order from quantum disorders (OFQD) phenomenon is well-known and ubiquitous
in particle physics and frustrated magnetic systems.
Typically, OFQD transfers a spurious Goldstone mode into a pseudo-Goldstone mode with a tiny gap.
Here, we report an opposite phenomenon:
OFQD transfers a spurious quadratic mode into a true linear Goldstone mode with a very small velocity
(named slow-Goldstone mode).
This new phenomenon is demonstrated in an interacting bosonic system subjected to an Abelian flux.
We develop a new and systematic OFQD analysis to determine
the true quantum ground state and the whole excitation spectrum.
In the weak-coupling limit,
the superfluid ground state has a 4-sublattice 90$^{\circ}$ coplanar spin structure,
which supports 4 linear Goldstone modes with 3 different velocities.
One of which is generated by the OFQD is much softer than the other 3 Goldstone modes,
so it can be easily detected in the cold atom or photonic experiments.
In the strong-coupling limit,
the ferromagnetic Mott ground state with a true quadratic Goldstone mode.
We speculate that there could be some topological phases intervening between the two symmetry broken states.
These novel phenomena may be observed in the current cold-atom or photonic experiments
subjected to an Abelian flux at the weak coupling limit where the heatings may be well under control.
Possible connections to  Coleman-Weinberg potential in particle physics, $ 1/N $ expansion of Sachdev-Ye-Kitaev models
and zero temperature quantum black hole entropy are outlined.}

\begin{document}
\maketitle
\flushbottom

\section{Introduction}
The Goldstone theorem states \cite{GS1,GS2,GS3}
that if a system has a continuous global symmetry $G$
which is spontaneously broken by its ground state to a smaller symmetry group $H$,
then 1) the symmetry broken ground state must support gapless modes,
2) the number of which is just the difference of the number of generators of $G$ and $H$.
These gapless modes with linear dispersions are called Goldstone modes.
The Goldstone theorem has had tremendous applications in essentially all the branches of physics
such as string theory/quantum gravity \cite{Stringbook,Stringbook2},
especially the gapless reparametrization mode in the Sachdev-Ye-Kitaev (SYK) models
leading to its maximal chaos tying that of a quantum black hole \cite{SY,Kit,Mald},
particle physics \cite{QFT}, condensed matter systems \cite{qaf1,qaf2}, cold-atoms/quantum optics
\cite{u1largen,gold,strongED}
and quantum information science \cite{QIbook}.
The number of Goldstone modes can be counted just from the symmetry breaking analysis.
However, it remains a challenge to evaluate their velocities on any specific Hamiltonian.

On the other hand, the order from quantum disorder phenomena (OFQD)
is quite common in particle physics \cite{CWpotential,CW2,weinberg_1996}
and frustrated magnetic systems \cite{subir,gan,frusrev,sachdev,Sachdev1991,
Balents2010,FrustratedBook}.
It says that there are infinitely many degenerate ground states 
at the classical level due to a spurious continuous symmetry.
There may also be an associated spurious Goldstone mode resulting from the breaking of such a spurious continuous symmetry.
However, quantum fluctuations can pick up quantum ground states from this classically degenerate manifold and generate a small gap to the spurious Goldstone mode, therefore transfer it into a pseudo-Goldstone mode.
Here we discover a completely opposite new phenomenon: the OFQD generates a true Goldstone mode with a very small velocity (named slow-Goldstone mode) which can be observed in the current spinor cold atoms
\cite{stagg1,stagg2-1,stagg2-2,stagg2-3,uniform1,uniform2,uniform3,newexp,newexp2,haldane}
or photonic experiments \cite{quater1,quater2,quater3,quater4} subjected to an Abelian $ \alpha $ flux in the weak coupling regime.

  The Abelian flux provides a frustrating source even in a square lattice 
  which is entirely different from the geometric frustrations
  which only happen in frustrated lattices.
  For simplicity and practical relevance to the cold atom or photon experiments,
  we focus on the most frustrated case with $ \alpha=\pi $ flux.
  We find that in the weak interaction limit, the $ \pi $ flux sparks dramatic interplay between the superfluid (SF)
  and the (pseudo-)spin of spinor bosons, which leads to a frustrated superfluid even in a bipartite lattice.
  We develop a new and systematic {\sl two-step} ``order from quantum disorder'' (OFQD) analysis to determine
  not only the ground state, but also the whole excitation spectrum of such a frustrated superfluid.
  In the first step, by identifying a classically degenerate family of states upto the exact spin $ SU(2)_s $ symmetry,
  and generalizing Bogoliubov theory to multi-components on a lattice, we find the quantum ground state is a superfluid state with
  the spin-orbital structure of a 4-sublattice $ 90^{\circ} $ coplanar state in a square lattice.
  It not only breaks the charge $ U(1)_c $ symmetry, but also  the spin $ SU(2)_s $ symmetry completely.
  The Goldstone theorem dictates that there should be 4 linear gapless modes.
  However, we find only 3 linear gapless modes, plus 1 quadratic mode in the long wavelength limit,
  in contradiction with the Goldstone theorem. Then we  devise the second step to
  incorporate the effective potential generated by the OFQD into the Hamiltonian.
  It transfers the quadratic mode into a soft linear gapless mode which
  is the Goldstone mode generated by the OFQD, but the other three linear modes  remain intact,
  in agreement with the exact results from the Goldstone theorem.
  Because the Goldstone mode generated by OFQD is much softer than the other three ones in the weak coupling, so it can be easily
  identified in the spinor cold atom or photonic experiments by various currently available detection methods.
  At the leading order of the strong coupling expansion, we find a ferromagnetic (FM) Mott state, now with a true quadratic FM mode.
  Due to the dramatically different nature of the two states in the weak and strong coupling, we
  speculate there could be a topological quantum spin liquid intervening between
  the weak coupling frustrated SF and the strong coupling FM Mott state.
  Because the heating effects can be easily suppressed at weak couplings,
  so this novel phenomenon due to OFQD at weak couplings can be easily observed in recent cold atom (or photonic) experiments
  generating Abelian flux in an optical lattice (or in a microwave cavity array).
  Several perspectives on Coleman-Weinberg potential in particle physics, $ 1/N $ expansion of Sachdev-Ye-Kitaev, and the quantum fluctuations effects at zero temperature on Bekenstein-Hawking entropy  are  outlined.

\section{The model and two-steps order from quantum disorder analysis}
\begin{figure}
\centering
\includegraphics[width=.68\linewidth]{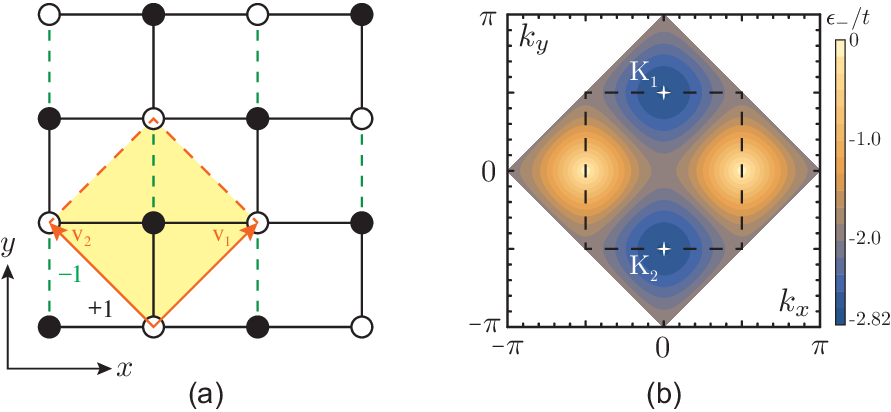}
\caption{
(a) The Z$_2$ gauge used in the main text with the real 
and spin-independent hopping $-te^{iA_{ij}}=\pm t$,
which leads to two sites per unit cell. 
The frustrated bond with $e^{iA_{ij}}=-1$ is denoted by a dashed line.
(b) In the Z$_2$ gauge, there are two (doubly) degenerate minima at the two momentum $\mathbf{K}_1=(0,\pi/2)$
and $\mathbf{K}_2=(0,-\pi/2)$. The diamond shape is the Brillouin zone (BZ). The inner dashed line is 1/2 of the BZ. The full square is twice the BZ (2BZ).}
\label{fig:frog}
\end{figure}

We consider a pseudo-spin-$1/2$  Boson-Hubbard model in a $\pi$-flux on the square lattice described by:
\begin{align}
    \mathcal{H}=
	-t \!\sum_{\langle ij\rangle,\sigma} (e^{i A_{ij} }
	 b_{i\sigma}^\dagger b_{j\sigma}\!+h.c.)
	+\frac{U}{2}\!\sum_{i} n^2_{i}
	-\mu\!\sum_{i} n_{i}\>,
\label{piflux}
\end{align}
where $\langle ij\rangle$ indicates the nearest-neighbor sites of a square lattice,
$b_{i\sigma}$ ($b_{i\sigma}^\dagger$) is the boson annihilation (creation) operators at site $i$ with a spin $\sigma$,
$A_{ij}$ is the static gauge potential on the link [see Fig.\ref{fig:frog}(a)],
$n_i=n_{i\uparrow}+n_{i\downarrow}$ is the total number of bosons for site $i$,
and $\mu$ is a chemical potential.
The second term with $U>0$ represents a repulsive on-site Bose-Hubbard interaction, 
which is invariant under spin $SU(2)$ symmetry.
As a result, the total Hamiltonian enjoys $U(1)_c \times SU(2)_s $ symmetry.

\subsection{Order from quantum disorder mechanism to determine the quantum ground state: step 1 }

In the weak coupling limit $ U/t \ll 1 $,
we will first diagonalize the non-interacting Hamiltonian
and then treat the weak interaction perturbatively by generalizing Bogoliubov theory to two components.
The gauge is shown in Fig.\ref{fig:frog}(a) leads to two sublattices $ A $ and $ B $.
The diagonalization in this gauge leads to single-particle spectrum 
$\epsilon_{\pm,\sigma}(\mathbf{k})=\pm2t\sqrt{\cos^2 k_x+\sin^2 k_y}$,
which develops two energy minima at momentum 
$\mathbf{K}_1=(0,\pi/2)$ and $\mathbf{K}_2=(0,-\pi/2)$
shown in Fig.\ref{fig:frog}(b) with the corresponding two-sublattice eigen-vectors:
\begin{align}
    \eta_1
     =\frac{1}{\sqrt{2}}
    \begin{pmatrix}
	    e^{-i\frac{\pi}{4}}\\
	    1\\
    \end{pmatrix},\quad
    \eta_2
    =\frac{1}{\sqrt{2}}
    \begin{pmatrix}
	    e^{i\frac{\pi}{4}}\\
	    1\\
    \end{pmatrix}\>.
\label{eta12}
\end{align}

At the weak coupling,
the ground-state is a superfluid state,
and all bosons will condense into the lowest single-particle energy eigenstate.
In the presence of two degenerated two degenerate energy minima
with momentum $\mathbf{K}_1=-\mathbf{K}_2=\mathbf{K}$,
we can construct a condensate wave-function as
\begin{align}
    \Psi_\mathbf{r}=\sqrt{2n}e^{i\chi}
	[(\eta_1\otimes \mathbf{z}_1)e^{-i\mathbf{K}\cdot\mathbf{r}}
	+(\eta_2\otimes \mathbf{z}_2)e^{+i\mathbf{K}\cdot\mathbf{r}}]
\label{weakpara}
\end{align}
where $\Psi_\mathbf{r}
=(b_{\mathbf{r},A,\uparrow},b_{\mathbf{r},A,\downarrow},
b_{\mathbf{r},B,\uparrow},b_{\mathbf{r},B,\downarrow})^\intercal$
is a four-component spinor,
$n$ is condensate density,
$\eta_{1,2}$ are sublattice eigen-vectors defined in Eq.\eqref{eta12},
and $\mathbf{z}_i=(z_{i\uparrow},z_{i\downarrow})^\intercal$ with $i=1,2$ 
are two-component non-normalized spinors
satisfying $ |\mathbf{z}_1|^2 + |\mathbf{z}_2|^2 =1 $.
From the construction, $\Psi_\mathbf{r}$ with arbitrary $\mathbf{z}_{1,2}$
automatically minimizes Kinetic energy,
but it gives different Bose-Hubbard interaction energy density 
\begin{align}
	E_\text{int}=\frac{n^2U}{2}(1+2|\mathbf{z}_1^\dagger \mathbf{z}_2|^2),
\label{E_int}
\end{align}
which suggests the mean-field ground-state manifold is described by the condition 
$\mathbf{z}_1^\dagger \mathbf{z}_2=0$.

\begin{figure}[tbhp]
\centering
\includegraphics[width=\linewidth]{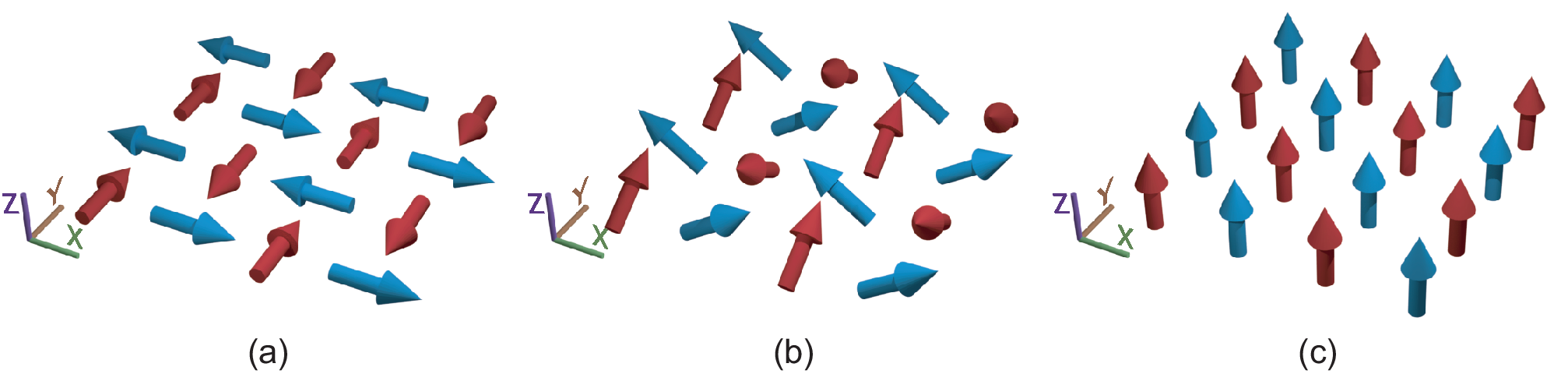}
\caption{
The spin structure of the superfluid (SF) states in the classically  $ U(1) $ degenerate manifold at several representative  $ \phi $.
(a) $\phi=\pi/4$ (coplanar), (b) $\phi=\pi/10$ (non-coplanar), (c) $\phi=0$ (collinear) 
ferromagnetism (FM), respectively.
The order from the quantum disorder  (OFQD) effect selects $\phi=\pi/4$ as the true quantum ground state,
which has the 4-sublattice coplanar $90^\circ$ frustrated spin structure given by  Eq.\eqref{state0spin}.
If changing the sign of the OFQD term, then it becomes the FM SF state (See Ref.\cite{QSLown}).  }
\label{classical}
\end{figure}

At the mean-field level, the condition $\mathbf{z}_1^\dagger \mathbf{z}_2=0$ 
leads to a degenerate family of ground-state,
which contains not only the exact $ U(1)_c \times SU(2)_s $ symmetries 
but also a spurious $ U(1) $ symmetry:
\begin{align}
    \Psi_0= \sqrt{2n_0} e^{i\chi}
	& [(\eta_1\otimes \mathbf{z}_{1}^{(0)})e^{-i\mathbf{K}\cdot\mathbf{r}}
	+(\eta_2\otimes \mathbf{z}_{2}^{(0)})e^{+i\mathbf{K}\cdot\mathbf{r}}]\>,
\label{state2}
\end{align}
where we fixed two spinors to be orthogonal
\begin{align}
	\mathbf{z}_{1}^{(0)}
	=e^{-i\alpha} \cos\phi
	\begin{pmatrix}
	    e^{-i\gamma/2}\cos\beta/2\\
	    e^{i\gamma/2}\sin\beta/2\\
	\end{pmatrix},\quad
	\mathbf{z}_{2}^{(0)}=e^{+i\alpha}\sin\phi
	\begin{pmatrix}
	    -e^{-i\gamma/2}\sin\beta/2\\
	    e^{i\gamma/2}\cos\beta/2\\
	\end{pmatrix}
\end{align}
The exact $ U(1)_c $ symmetry tunes the global phase $ \chi $.
The exact $ SU(2)_s $ symmetry tunes the three spin angles $(\alpha,\beta,\gamma)$,
while the spurious $ U(1) $ symmetry tunes the angle $\phi$, 
which need to be determined by the ``order from quantum disorder'' mechanism. 
Three representative classically degenerate states are given in
Fig.\ref{classical}.

Using the exact $ U(1)_c $ symmetry and the exact $ SU(2)_s $ symmetry, 
we set $ \chi =0 $ and align the spinor along the spin-$Z$ direction.
Equation \eqref{state2} can be simplified to
\begin{align}
    \Psi_0  =\sqrt{\frac{N_0}{N_s}}
	&\left [
	    \cos\phi
	    \begin{pmatrix}
	    e^{-i\frac{\pi}{4}}\\1\\
	    \end{pmatrix}
	    \otimes
	    \begin{pmatrix}
	    1\\0\\
	    \end{pmatrix}
	    e^{-i\mathbf{K}\cdot\mathbf{r}}  \right.  
        \left. + \sin\phi
	    \begin{pmatrix}
	    e^{i\frac{\pi}{4}}\\1\\
	    \end{pmatrix}
	    \otimes
	    \begin{pmatrix}
	    0\\1\\
	    \end{pmatrix}
	    e^{+i\mathbf{K}\cdot\mathbf{r}}
	\right]\>,
\label{state0}
\end{align}
where $N_s$ stand for the total number of lattice sites,
and $N_0$ stand for the total number of condensed atoms,
and $n_0=N_0/N_s$ is the condensate density.
The density and spin of the state Eq.\eqref{state0} can be evaluated as:
\begin{align}
	\langle\Psi_0^s|\sigma_0|\Psi_0^s\rangle&=n_0;   \nonumber \\
	\langle\Psi_0^s|\sigma_x|\Psi_0^s\rangle&
	    =n_0\cos(2\mathbf{K}\cdot\mathbf{r})\sin2\phi,~ 0 \quad
	    \text{ $s \in B, A$ ;}    \nonumber  \\
	\langle\Psi_0^s|\sigma_y|\Psi_0^s\rangle&
	    =n_0\cos(2\mathbf{K}\cdot\mathbf{r})\sin2\phi,~0  \quad
	    \text{ $s \in A, B$ ;}     \nonumber  \\
	\langle\Psi_0^s|\sigma_z|\Psi_0^s\rangle&=n_0\cos2\phi.
\end{align}
where $\mathbf{r}$ stands for the unit cell coordinate in Fig.\ref{fig:frog}(a)
and $s=A,B$ stands for the two sublattices,  
i.e. $\mathbf{r}_{B}=\mathbf{r}_{A}+\hat{y}$.
%
In terms of the coordinate-$i$ in the square lattice, the state has the spin structure
$ \mathbf{S}_i=(S_i^x,S_i^y,S_i^z) $:
\begin{align}
	\mathbf{S}_i
	=\frac{n_0}{2}\left(
	    \frac{(-1)^{i_x}-(-1)^{i_y}}{2}\sin2\phi,
	    \frac{(-1)^{i_x}+(-1)^{i_y}}{2}\sin2\phi,
    \cos2\phi  \right)\>.
\label{Sphi}
\end{align}
Setting $ \phi=0, \pi/10, \pi/4 $ lead to the three typical states in Fig.\ref{classical}

By writing the Bose field as a sum of condensation plus a small quantum fluctuation
$\Psi= \Psi_0 + \psi $, one can perform an expansion of original Hamiltonian
$\mathcal{H}=\mathcal{H}^{(0)}+\mathcal{H}^{(1)}+\mathcal{H}^{(2)}+\cdots$, 
where $\cdots$ means high order terms.
The first term in the expansion is
$\mathcal{H}^{(0)}=E_0=(-2\sqrt{2}t-\mu+\frac{1}{2}n_0U)N_0$,
which is the mean-field ground-state energy,
then setting $\mathcal{H}^{(1)}=0$  determines chemical potential
 $ \mu=-2\sqrt{2}t+n_0U $ and  $ E_0=-\frac{1}{2}n_0U N_0 $.
    Diagonalizing $ \mathcal{H}^{(2)} $
    by a $8\times8$ Bogoliubov transformation leads to:
\begin{align}
    \mathcal{H}^{(2)}
	=E^{(2)}_0	+\sum_\mathbf{q}\sum_{l=1}^4\Omega_l(\mathbf{q};\phi)
	\left(\beta_{l,\mathbf{q}}^\dagger\beta_{l,\mathbf{q}}+\frac{1}{2}\right)
\label{h20}
\end{align}
where $ E^{(2)}_0=-N_s(2\sqrt{2}t+n_0U/2) $
and $ \mathbf{q}\in\{(q_x,q_y):|q_x+q_y|\leq\pi\} $ is the diamond Brillouin zone (BZ) in Fig.\ref{fig:frog}(b). 
\footnote{
In principle, due to the 2 sublattices, 2 valleys, 2 spin, and 2 particle-hole structures, the $ 16 \times 16 $ Bogoliubov transformation leads to 8 bands in the 1/2 BZ in Fig.\ref{fig:frog}(a). 
However, we find that we can shift 4 of the 8 bands to make the 4 bands $ \Omega_l, l=1,2,3,4 $ in the (diamond) BZ in Fig.\ref{fig:frog}(b).}
The quantum corrected ground state energy is:
\begin{align}
    E_\text{GS}(\phi)
	=E_{t0}
	+\frac{1}{2}\sum_\mathbf{q}\sum_{l=1}^4 \Omega_l(\mathbf{q};\phi)
\end{align}
where $ E_{t0}=E_0+ E^{(2)}_0 $.

\begin{figure}
\centering
\includegraphics[width=0.75\linewidth]{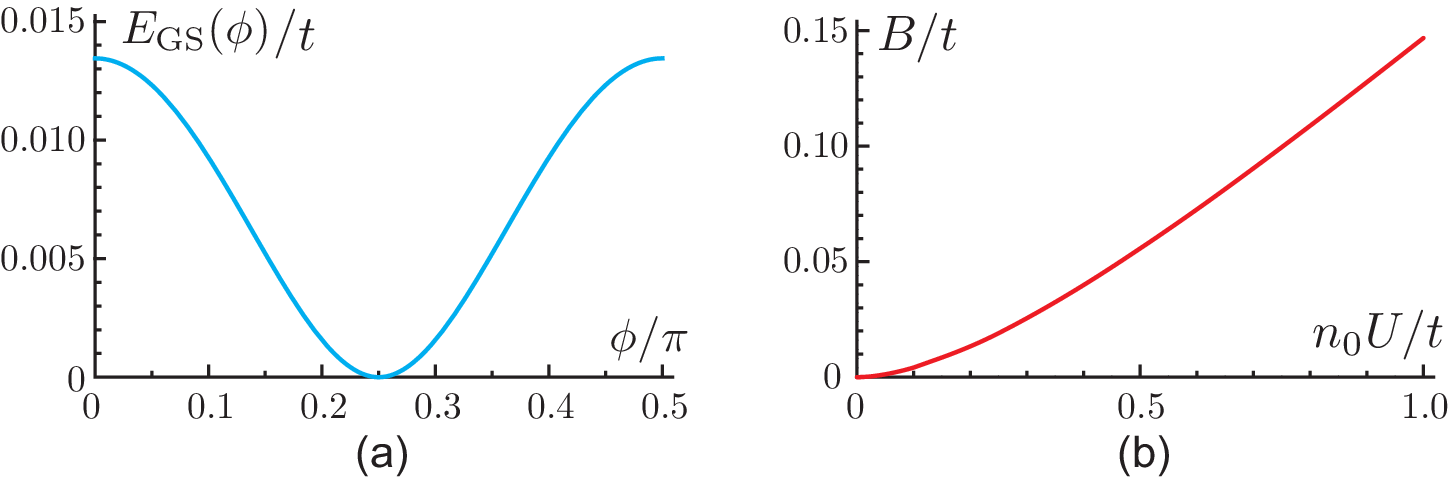}
\caption{
(a)The quantum ground-state energy $E_\text{GS}( \phi) $ has a minimum at $\phi=\pi/4$.
	Expanding around the minimum leads to the coefficient $ B $ in (b),
which is a monotonically increasing function of the interaction strength.
In order to clearly see the OFQD, we subtract $E_\text{GS}$ by it minimum $\min_\phi E_\text{GS}$.
The parameters are $n_0U=t=1$.}
\label{degree90}
\end{figure}

The numerical result of $ E_\text{GS}(\phi) $ as a function of $\phi$ in Fig.\ref{degree90}(a) shows that $E_\text{GS}$ reaches its minima at $\phi=\pi/4$.
Thus, we can expand $ E_\text{GS}(\phi) $ around its minimum $\phi=\pi/4+\delta\phi$ as
\begin{align}
    E_\text{GS}(\phi,n_0U)
	=E_1 + \frac{B}{2}(\delta\phi)^2+\cdots
\label{B2}
\end{align}
where $ E_1=E_\text{GS}(\phi=\pi/4,n_0U) $ and  the coefficient $B$ is obtained as
\begin{align}
	B=(n_0U)^2\sum_{\mathbf{q}\in\text{BZ}}\sum_{n=1}^{4}
	\frac{2^8t^4(\cos q_x-\cos q_y)^2\Omega_n(\mathbf{q})}
	{\prod_{m\neq n}[\Omega_n(\mathbf{q})^2-\Omega_m(\mathbf{q})^2]}	
\end{align}
where $\Omega_n= \Omega_n(\mathbf{q}, \phi=\pi/4 )$. We find that  $ B=c\frac{(n_0U)^2}{t}  $ where $ c \sim a + b \ln (U/t)$ as shown in Fig.\ref{degree90}(b).

So the ``order by disorder'' mechanism selects $\phi=\pi/4$  among the classically degenerate family.
Setting $ \phi=\pi/4 $ into Eq.\eqref{Sphi} leads to the associated spin structure:
\begin{align}
	\mathbf{S}_i=\frac{1}{4}[(-1)^{i_x}(1,1,0)+(-1)^{i_y}(-1,1,0)]
\label{state0spin}
\end{align}
   which is a 4-sublattice $ 90^{\circ} $ spin structure in Fig.\ref{classical}(a), in comparison with the 3-sublattice $ 120^{\circ} $
   frustrated spin structure in a triangular lattice \cite{gan,frusrev,sachdev}.


\subsection{Excitation spectrum above the $ 90^{\circ} $ frustrated  quantum ground state: a partial contradiction to the Goldstone theorem }
  Setting $\phi=\pi/4$ into the Bogliubov Hamiltonian Eq.\eqref{h20}, 
  we find that
  $ \Omega_1(\mathbf{q}) $ contains one linear mode at  $ (0,0) $  ( called $ \omega_1 $  ) and
  another at $ (\pi,0) $ ( called $ \omega_2 $  ).
  $ \Omega_2(\mathbf{q}) $ contains one quadratic  mode at  $ (0,0) $  ( called $ \omega_3 $  ) and
  another at $ (\pi,0) $ ( called $ \omega_4 $ ). Their expressions can be simplified  if we introduce:
\begin{align}
    \omega_\pm(\mathbf{q})
	=\sqrt{A_\mathbf{q}+B_\mathbf{q}\pm\sqrt{4A_\mathbf{q}B_\mathbf{q}+C_\mathbf{q}^2}}\>,
\label{wpm}
\end{align}
where we define
\begin{eqnarray}
    A_\mathbf{q}
	& = &4t^2(\cos^2 q_x\!+\cos^2 q_y)\!-\!\sqrt{2}tn_0U(\cos q_x\!+\cos q_y),  \\
    B_\mathbf{q}
	& = & 2t(4t+\sqrt{2}n_0U),    \\
    C_\mathbf{q}
	& = & \sqrt{2}tn_0U(4\!+\cos q_x\!+\cos q_y)\>,
\label{abc}
\end{eqnarray}
where $ \mathbf{q} $ is defined in the 2BZ in Fig.\ref{fig:frog}(b).
 In the long wavelength limit $ \mathbf{q} \to 0 $, the 4 gapless modes can be written as:
$\omega_1  = \omega_-(\mathbf{q}) = \sqrt{\sqrt{2}tn_0U}q$,
$\omega_2= \omega_-( \mathbf{K}_2 +\mathbf{q})=\omega_4= \omega_-( \mathbf{K}_4 +\mathbf{q}) = \sqrt{\frac{4t^2n_0U}{4\sqrt{2}t+n_0U}}q$,
$\omega_3  = \omega_-(\mathbf{K}_3 +\mathbf{q})= \frac{t}{\sqrt{2}}q^2$ 
where $ \mathbf{K}_2=(\pi,0), \mathbf{K}_4=(0, \pi), \mathbf{K}_3=(\pi,\pi) $
\footnote{Note that these $\mathbf{K}_{1,2}$ defined in the 2BZ are different from
its original definitions in the (diamond)  BZ in Fig.\ref{fig:frog}(b). }.
Here the degeneracy  $ \omega_2=\omega_4 $ is dictated by the 
$ [C_4 \times C_4]_\text{Diagonal} $ symmetry of the $ 90^{\circ} $ state in Fig.\ref{classical}(a). 
One can see that the three linear modes $ \omega_1, \omega_2  = \omega_4 $ are proportional to $ \sqrt{ n_0 U t} $, 
while the quadratic mode  $ \omega_3 $ is independent of the interaction $ U $. 
In fact, it is identical to the free particle dispersion.
On the other hand, the $ 90^{\circ} $  coplanar state in Fig.\ref{classical}(a)
  breaks both the $ U(1)_c $ and completely the spin $ SU(2) $ symmetry, so it should lead to
  1+3 linear gapless modes instead of 3 linear and 1 quadratic mode.
  Therefore the quadratic mode  $ \omega_3 $ must be an artifact of the Bogliubov calculation to this order.
  In the following, we show that this in-consistence can be completely resolved
  by involving the order from quantum disorder mechanism.

\subsection{The slow-Goldstone mode generated by the OFQD mechanism: 
step 2, a complete path integral formalism of the OFQD }

In the geometrically frustrated quantum spin systems, an ``order from quantum
disorder'' analysis \cite{gan,frusrev,sachdev} was developed to calculate the gap at 
$ \mathbf{k}=0 $.
When applying this analysis to Eq.\eqref{B2},
we find that due to the absence of the conjugate $ A $ term 
(as dictated by a $ U(1) $ subgroup of the spin $ SU(2) $ symmetry),
there is still no gap at $ \mathbf{q}=0 $ generated to the quadratic $\omega_3$ mode.
So the previous analysis in \cite{gan,frusrev,sachdev} can not be used to find any useful information to the present problem.
Here, we develop the second step of the systematic ``order from disorder analysis'' to go beyond that developed  in \cite{gan,frusrev,sachdev} to compute the contributions to the spectrum from Eq.\eqref{B2}.
There are two alternative approaches, 
one is the canonical quantization approach to be presented in detail in appendix A,
the other is the boson coherent state path integral approach presented here.
The latter
 is more systematical and powerful than the canonical quantization approach.
 Most importantly, it can also be used to capture the non-linear effects, so it is essentially a non-linear sigma model
 approach that can be applied directly to study finite temperature effects.
 The physical picture is also more transparent in the polar coordinates,
 especially in identifying the slow-Goldstone mode generated by OFQD.

To capture low-energy excitation above the ground-state,
we re-examine Eq.\eqref{weakpara} and Eq.\eqref{E_int},
and study quantum fluctuations rather than just look at the ground-state.
In the $S_z$ basis, it is naturally to introduce a parametrization of $\mathbf{z}_{1,2}$
\begin{align}
    \mathbf{z}_1=
	e^{-i\theta}\cos\phi
	\begin{pmatrix}
	    e^{-i\alpha_1/2}\cos(\beta_1/2)\\
	    e^{+i\alpha_1/2}\sin(\beta_1/2)\\
	\end{pmatrix}, \quad   
    \mathbf{z}_2=
	e^{+i\theta}\sin\phi
	\begin{pmatrix}
	    e^{-i\alpha_2/2}\cos(\beta_2/2)\\
	    e^{+i\alpha_2/2}\sin(\beta_2/2)\\
	\end{pmatrix},
\label{z1z2phi}
\end{align}
then interaction energy density becomes
\begin{align}
    E_\text{int}\!=\!
	\frac{n^2U}{2}
	\Big\{
		1\!+\!\frac{1}{2}\sin^2(2\phi)
	    \Big[\cos^2\big(\frac{\alpha_1\!-\!\alpha_2}{2}\big)
	    \cos^2\big(\frac{\beta_1\!-\!\beta_2}{2}\big)
	    +\sin^2\big(\frac{\alpha_1\!-\!\alpha_2}{2}\big)
	    \cos^2\big(\frac{\beta_1\!+\!\beta_2}{2}\big)\Big]
	\Big\}
\end{align}
and minimization condition can be expressed as 
$\alpha_1-\alpha_2=0$ and $\beta_1-\beta_2=\pm\pi$.

In the mean-field ground-state manifold,
the interaction energy is independent of choice of 
$\chi$, $\theta$, $\alpha_1+\alpha_2$, $\beta_1+\beta_2$ and $\phi$,
where the first one is related to U(1)$_c$ symmetry,
the second to fourth ones are related to SU(2)$_s$ symmetry, 
but the last one $\phi$ is due to spurious symmetry.
The mean-field ground-state in Eq.\eqref{state2}
corresponds to choose 
$\theta=\alpha$, $\alpha_1=-\alpha_2=\gamma$, $\beta_1=\beta_2-\pi/2=\beta$,
and then fix $\chi=\alpha=\beta=\gamma=0$ in Eq.\eqref{state0}.
Here, we prefer to choose a different but equivalent saddle point solution
\begin{align}
    n=n_0,\quad
    \chi=0,\quad
    \alpha_1=\alpha_2=0,\quad
    \beta_1=\beta_2+\pi=\pi/2\>.
\label{saddle11}
\end{align}
The corresponding $\mathbf{z}$ spinors are:
  $  \mathbf{z}_1=
	(\begin{smallmatrix}
	    1/2\\1/2\\
	\end{smallmatrix}),
    \mathbf{z}_2=
	(\begin{smallmatrix}
	    1/2\\-1/2\\
	\end{smallmatrix}) $, 
which are related to those in Eq.\eqref{state0} by a $ SU(2)_s $ rotation.
The former is in the $ \sigma_x $ representation, 
the latter is in the $ \sigma_z $ representation
which may be singular in the polar coordinate.

The Bose-Hubbard interaction can be expanded around the saddle point as
\begin{align}
    \mathcal{H}_\text{int}\!
	=&\frac{Un_0^2}{2}N_s
	+\!\sum_\mathbf{r} 2Un_0\delta n
	+\!\sum_\mathbf{r}U(\delta n)^2 
	+\!\sum_\mathbf{r}\frac{1}{8}U n_0^2[
		(\delta\beta_{1}\!-\!\delta\beta_{2})^2
		+(\delta\alpha_{1}\!-\!\delta\alpha_{2})^2]\>.
\end{align}
Similarly, the Kinetic energy term becomes
\begin{align}
	\mathcal{H}_\text{kin}
	&=\sum_\mathbf{k} \Psi_\mathbf{k}^\dagger(\epsilon_{-}-\mu)\Psi_\mathbf{k}
	\nonumber\\
	&\approx-2\sqrt{2}tn_0 N_s-\mu n_0 N_s 
	-2\sum_\mathbf{r} (2\sqrt{2}tn_0 N_s+\mu)\delta n
	-\sqrt{2}n_0t\sum_\mathbf{r} \Psi^\dagger\nabla^2\Psi\>.
\end{align}
Similar to the Bogoliubov theory, 
the linear term eventually got cancelled by $\mu=-2\sqrt{2}t+n_0U$.
After including the boson Berry phase $ \Psi^\dagger\partial_\tau \Psi $  term 
and order-from-disorder generated potential,
we form a complete Lagrangian density
\begin{align}
    \mathcal{L}_\text{SF}
	&=i\delta n\partial_\tau\delta\chi
	+\frac{i}{4}n_0
	(8\delta\phi\partial_\tau\delta\theta
	+\delta\beta_1\partial_\tau\delta\alpha_1
	-\delta\beta_2\partial_\tau\delta\alpha_2)\nonumber\\
	&-\frac{n_0t}{\sqrt{2}}
	[\frac{1}{4n_0^2}(\nabla\delta n)^2+(\nabla\delta\chi)^2
	+(\nabla\delta\theta)^2
	+(\nabla\delta\phi)^2]\nonumber\\
	&-\frac{n_0t}{8\sqrt{2}}
	[(\nabla\delta\alpha_1)^2
	+(\nabla\delta\alpha_2)^2
	+(\nabla\delta\beta_1)^2
	+(\nabla\delta\beta_2)^2]\nonumber\\
	&+\frac{1}{2}U(\delta n)^2
	+\frac{1}{16}n_0^2U[
	(\delta\beta_1-\delta\beta_2)^2
	+(\delta\alpha_1-\delta\alpha_2)^2]\nonumber\\
	&+\frac{B}{2}(\delta\phi)^2\>,
\end{align}
where the last term is the order-from-disorder generated potential.
The Lagrangian density contains four conjugate pairs:
$(\delta n,\delta\chi)$, $(\delta\phi,\delta\theta)$,
$(\delta\beta_1,\delta\alpha_1)$, $(\delta\beta_2,\delta\alpha_2)$.
The conjugate pair $(\delta n,\delta\chi)$ 
is density fluctuations and global phase fluctuations, 
which gives superfluid sound mode;
in contrast,
the other three pairs contribute to the spin fluctuations,
which give the three spin modes.
The conjugate pair $(\delta\phi,\delta\theta)$ is associated with the spurious symmetry.
Before considering the order from quantum disorder effect,
$(\delta\phi,\delta\theta)$ behaves just like two transverse components of the magnetization in a ferromagnet, 
which results in a ``spurious'' quadratic dispersion $\omega\sim q^2$. 
However, after including the order from quantum disorder effect,
$\delta\phi$ acquires a small finite OfD generated potential,
thus it modifies the quadratic dispersion to a true linear one $\omega\sim q$.

\begin{figure}
\centering
\includegraphics[width=0.8\linewidth]{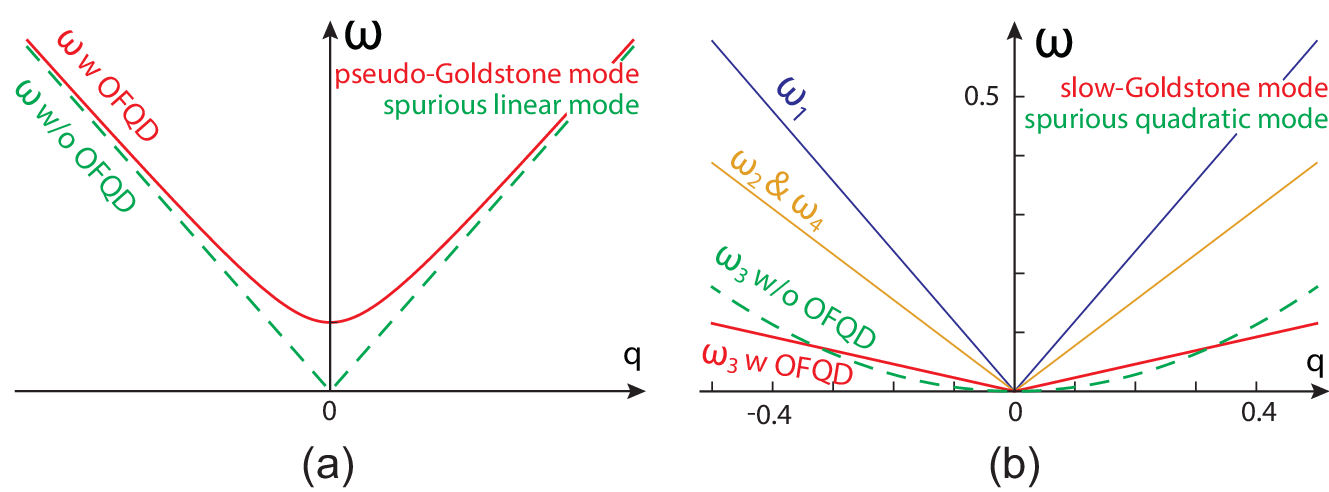}
\caption{
The pseudo-Goldstone mode discovered in previous works vs slow-Goldstone mode discovered in this work:
(a)Typical OFQD generated pseudo-Goldstone mode.
The OFQD transfers a spurious linear mode ($\omega$ w/o OFQD) into
a pseudo-Goldstone mode ($\omega$ w OFQD) with a tiny gap.
(b)The OFQD generated slow-Goldstone mode ($\omega_3$)
and the other 3 conventional Goldstone modes ($\omega_{1,2,4}$).
The OFQD transfers a spurious quadratic mode ($\omega_3$ w/o OFQD c.f. Eq.\eqref{wpm}) into
a slow-Goldstone mode ($\omega_3$ w OFQD c.f. Eq.\eqref{fourl}) with a very small velocity.}
\label{GoldstoneFig}
\end{figure}

Finally, In the long-wave length limit,
these four conjugate pairs give 4 linear gapless modes:
\begin{eqnarray}
    \omega_1 & = & v_1 q=\sqrt{\sqrt{2}tn_0U}q,  \nonumber   \\
    \omega_2 & =  & \omega_4=v_2 q=\sqrt{\frac{4t^2n_0U}{4\sqrt{2}t+n_0U}}q \sim \omega_1/\sqrt{2}   \nonumber   \\
    \omega_3 & =  &  v_3 q=\sqrt{\frac{Bt}{2\sqrt{2}n_0}}q
\label{fourl}
\end{eqnarray}
where $q=|\mathbf{q}|=\sqrt{q_x^2+q_y^2}$.
 Especially, we find $ ( \theta, \phi ) $ conjugate pair in Eq.\eqref{z1z2phi}
 is the slow-Goldstone mode generated by OFQD.
 As a comparison, we illustrate the OFQD generated pseudo-Goldstone mode
 and OFQD generated slow-Goldstone mode in Fig.\ref{GoldstoneFig}(a) and (b),
 respectively.
  Using the fact $ B= c (n_0 U)^2/t $, its velocity $ v_3 \sim  n_0^{1/2} U $ vanishes in the $ U\to0 $ limit.
  In the weak coupling limit,  $ U/t \ll 1 $,  we have relation
\begin{equation}
  v_1>v_2=v_4 \sim v_1/\sqrt{2} \gg v_3\>.
\end{equation}
So the $ \omega_3 $ mode becomes linear only in a tiny regime near zero momentum, 
beyond which it recovers the quadratic free particle mode. 
The $ \omega_3 $ mode is the slow-Goldstone mode generated by the order from quantum disorder. 
Because the slow-Goldstone mode is much softer than the other three Goldstone modes, 
it can be easily identified in the spinor cold atom or photonic experiments.

\subsection{The condensate fraction and the quantum depletions}
 From Eq.\eqref{h20},\eqref{wpm},\eqref{abc}, we can evaluate the quantum depletion $ N_1= N-N_0 $:
\begin{align}
    N_1 &=\frac{1}{4}\sum_{\mathbf{q},s=\pm}
	[\frac{1}{\omega_s(\mathbf{q})}
	( 4\sqrt{2} t+n_0U              
	        +s\frac{2(4\sqrt{2} t+n_0U)A_\mathbf{q}+n_0UC_\mathbf{q}}
	       {\sqrt{\smash[b]{4A_\mathbf{q}B_\mathbf{q}+C_\mathbf{q}^2}}}  )-1 ]
\label{n1}
\end{align}
where sum over $\mathbf{q}$ is restricted to the 2BZ in Fig.\ref{fig:frog}(b).

\begin{figure}
\centering
\includegraphics[width=0.8\linewidth]{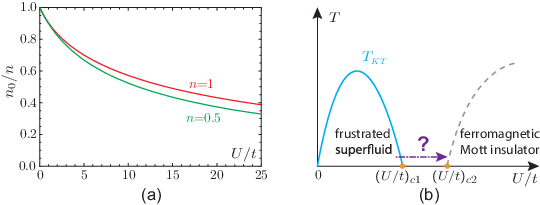}
\caption{ (a) The condensate fraction $n_0/n$ as a function of $U/t$ when filling factor $n=0.5$ and $ n=1 $.
(b) At $ T=0 $,  there could be some kind of quantum spin liquid (QSL) intervening
between the frustrated SF at weak coupling and  the FM SF at strong coupling. }
\label{condensate}
\end{figure}

In the cold atom experiments, the total density $ n $ is given. Then
the condensate density $n_0$ can be solved with a fixed values of $n$ and $U$ in Eq.\eqref{n1}.
Shown in Fig.\ref{condensate}(a) is the condensate fraction $n_0/n$ as a function of $n$ and $U$.
where one can see at $ U/t \ll 1 $, $ 1- n_0/n \ll 1 $ justifies the perturbation theory in the weak coupling limit.
As $ U/t $ gets larger, at integer fillings $ n $, the system remains in the SF state when $ U/t $ is below a critical value [Fig.\ref{condensate}(b)],
the ground state remains the 4-sublattice coplanar $ 90^{\circ} $ state in Fig.\ref{degree90}
  with the 4 gapless linear modes.  
However, their velocities in Eq.\eqref{fourl} may not be precise anymore,
when near $ (U/t)_c $.

\subsection{The strong coupling ferromagnetic Mott state}
In the strong coupling limit $ U/t \gg 1 $, at the integer fillings, 
the system gets into a Mott insulating state.
The second order perturbation in the strong coupling expansion in $ t^2/U $ lead to a spin-$n/2$ Ferromagnetic
Heisenberg model \cite{rh}:
\begin{align}
    \mathcal{H}_\text{spin}
	=-J\sum_{\langle ij\rangle}\mathbf{S}_i\cdot\mathbf{S}_j
\label{eq:effSpinH}
\end{align}
where $J=4t^2/U$. 
Note that the $0$ or $\pi$ flux makes no difference in the strong coupling limit to this order.


So the ground state in the strong coupling limit  is nothing but the FM Mott state.
The dispersion in the long wavelength limit is the well-known quadratic gapless FM mode 
\begin{equation}
    \omega_\text{FM}(k)=\frac{nt^2}{U}(k_x^2+k_y^2)\>,
\label{FMmode}
\end{equation}
which is due to the spontaneous symmetry breaking $ SU(2)_s \rightarrow U(1)_s $.
This is in sharp contrast to the symmetry breaking
$ U(1)_c \times SU(2)_s \rightarrow 1 $ 
leading to 4 linear Goldstone modes in the frustrated  SF phase in Eq.\eqref{fourl}.
Note that the FM state included in the classically  degenerate family of states at the weak coupling in Fig.\ref{classical}(c) is a SF state instead of a Mott state.

As shown in the previous paragraph, at weak coupling, 
there is a BEC$\to$superfluid at $ T=0 $, 
then the $ U(1)_c \times SU(2)_s \rightarrow 1 $ is completely broken, 
which leads to the 4 Goldstone modes, and one of which is the slow-Goldstone mode.
There is also an associated lattice symmetry breaking shown in Fig.\ref{classical}.
At any finite $T$, the continuous symmetries have to be restored, 
so only one of the 4 Goldstone modes: 
the superfluid one $ \omega_1 $ survives.
The finite  temperature phase transition at weak coupling could be just 
the Kosterlitz–Thouless (KT) transition [Fig.5(b)].
Of course, the FM order at strong coupling also disappears at any finite $ T $.
The outstanding problem remaining is to find what could be the intervening phase between the two limits in Fig.\ref{condensate}(b).
This will be addressed in the following work \cite{QSLown}.

\section{The implications on cold atom and photonic experiments}

  In fact, the problem of a particle moving in a lattice subject to an Abelian flux has a long history.
  The most original system is free electrons moving in a solid subject to a magnetic flux $ \alpha $
  through a unit cell which is called Hofstadter problem \cite{hh-1,hh-2,hh-3}. Unfortunately, it needs an astronomical
  large magnetic field to generate any appreciable
  magnetic flux $ \alpha $  through each unit cell of a solid, so it is difficult
  to realize the Hofstadter problem of free electrons in any solid  
  (However, for a recent progress in graphene, see \cite{BLgraphene}).
  Fortunately, there are recent experimental advances to generate
  effective magnetic flux $ \alpha $ through each unit cell of an optical lattice for cold atoms
  \cite{stagg1,stagg2-1,stagg2-2,stagg2-3,uniform1,uniform2,uniform3,newexp,newexp2,haldane}
  and that of a microwave cavity array for photons \cite{quater1,quater2,quater3,quater4}.
  Both the cold atom and the photonic systems lead to the bosonic analog of Hofstadter problem.
  Of course, in contrast to fermions, bosons are necessarily interacting,
  so one must incorporate interactions onto the problem.

On the other hand, it was well known the cold atom systems 
in an optical lattice suffer the heating problem.
Fortunately, it is still well under control with enough long lifetime at weak coupling, 
but gets worse as the coupling strength increase. 
So the frustrated SF at the weak couplings in Fig.\ref{condensate}
can be directly probed in the current cold atom experiments, 
but the FM may not (at this stage).
In this section, we only focus on the experimental detections in the weak coupling regime.

The cold atom condensation wave-function can be directly imaged 
through time-of-flight (TOF) images \cite{blochrmp}, 
which after a time $ t $ is given by:
\begin{equation}
     n( \mathbf{r} )= ( M/\hbar t )^3 f(\mathbf{k}) G(\mathbf{k})\>,
\end{equation}
where $ \mathbf{k}= M \mathbf{r}/\hbar t $,  
$ f(\mathbf{k})= | w(\mathbf{k}) |^2 $ is the form factor 
due to the Wannier state of the lowest Bloch band of the optical lattice, 
and $ G(\mathbf{k}) = \frac{1}{N_s} \sum_{i,j} e^{- \mathbf{k} \cdot ( \mathbf{r}_i- \mathbf{r}_j ) } \langle \Psi^{\dagger}_i \Psi_j \rangle $ is the equal time boson structure factor.
For a small condensate depletion shown in Fig.5(a)
$ \langle \Psi^{\dagger}_i \Psi_j \rangle \sim \langle \Psi^{\dagger}_{0i} \Psi_{0j}  \rangle$,
where $ \Psi^{\dagger}_{0i} $ is the condensate wave-function Eq.\eqref{state0} at $ \phi=\pi/4 $.
So the TOF can detect the quantum ground state wave-function directly.
The quantum depletion calculated in Eq.\eqref{n1} leads to a reduction in the magnitude of the condensation peaks at $ \mathbf{K} $ and $ -\mathbf{K} $ in Eq.\eqref{state0} and some broad backgrounds.

The 4-sublattice $ 90^{\circ} $ spin-orbital structure Eq.\eqref{state0spin} in Fig.\ref{degree90},
the 4 gapless modes in Eq.\eqref{fourl} and the nature of the transition in Fig.\ref{condensate}(b) can be precisely determined by dynamic or elastic, energy or momentum resolved, longitudinal or transverse Bragg spectroscopies \cite{braggbog-1,braggbog-2,braggbog-3,braggbog-4,braggbog-5,braggangle,braggeng,braggsingle-1,braggsingle-2,braggsingle-3,becbragg,bragg12-1,bragg12-2} in cold atoms and the site- and time-resolved spectroscopy \cite{quater1,quater2,quater3,quater4} in photonic systems.
As stressed below Eq.\eqref{fourl},  
the Goldstone mode $ \omega_3 $ generated by OFQD
is much softer than the other three Goldstone modes, 
so it can be easily distinguished by these experimental techniques.
Notably, both Goldstone and Higgs modes were detected near the two-dimensional superfluid/Mott insulator transition \cite{SS1} and in a supersolid cold atomic quantum gas \cite{SS2}.

\section{Discussion and conclusion}

The OFQD was first discovered in the geometrically frustrated magnets \cite{gan,frusrev,sachdev}.
However, so far, it was developed only to find the quantum ground state and then calculate the gap generated by the phenomena, but not the spectrum yet. 
Here, we developed a new and systematic two-step OFQD analysis to calculate
not only the mass gap but also the whole excitation spectrum. 
This systematic development, especially its second step, 
is vital to discover the slow-Goldstone mode generated by the OFQD.
The new method developed in this work can be transformed to compute the whole excitation spectrum in all the other frustrated systems such as geometric frustrated systems,
a system subject to an non-Abelain flux 
and particle physics.
The novel phenomenon of the slow-Goldstone mode generated by OFQD 
may also appear in these quantum systems.

As mentioned in the introduction, there may be also an associated spurious Goldstone mode resulting from the breaking of a spurious continuous symmetry. 
Then quantum fluctuations may generate a small gap to this spurious Goldstone mode
  and transfer it into a pseudo-Goldstone mode.
  In particle physics, there is a closely related phenomena called the  Coleman-Weinberg effective potential \cite{CWpotential}.
  For example, the pion with a light mass maybe just such a pseudo-Goldstone mode \cite{CW2,weinberg_1996}.
  The putative axion with a tiny mass maybe also such a pseudo-Goldstone mode \cite{axion,axion2,axion3} with the approximate
  global (Peccei-Quinn) $ U(1) $  symmetry corresponding to the spurious $ U(1) $ symmetry here.
  Here we discover a completely opposite new phenomenon: a true slow-Goldstone mode generated by the OFQD,
  which can be observed in the current spinor cold atom or photonic experiments subject to an Abelian $ \alpha= \pi $ flux.
  It is important to explore possible connections between the effective potential generated by the OFQD
  presented here and the Coleman-Weinberg potential in relativistic quantum field theory.
  It is also interesting to see if this opposite new phenomenon may also happen in particle physics.

It is instructive to look at the OFQD from the entropy point of view.
At the classical level, the spurious $ U(1) $ symmetry leads to a classically degenerate family of ground-states which, in turn, leads to an extensive ground state degeneracy, 
therefore an extensive $ T=0 $ entropy, 
clearly violating the third law of thermodynamics. 
However, as shown here, the OFQD spoils the spurious $ U(1) $ symmetry,
picks a unique quantum ground state,  
therefore leads to a vanishing  the $ T=0 $ entropy, 
consistent with the third law of thermodynamics. 
In the SYK models \cite{SY,Kit,Mald}, at the mean-field level $ N= \infty $,
there is also an extensive ground state degeneracy, 
therefore the extensive $ T=0 $ entropy $ s_0 N $ with $ s_0=0.2324... $, 
clearly violating the third law of thermodynamics. 
However, the $ 1/N $ expansion picks a unique quantum ground state, 
therefore leads to a vanishing the $ T=0 $ entropy consistent 
with the third law of thermodynamics.
It was argued that the $ T=0 $ entropy $ s_0 $ corresponds to the classical Bekenstein-Hawking (BH) entropy of a classical black hole in the $ AdS $ bulk.
It remains interesting to see if the quantum fluctuations will drive to classical BH entropy for an extreme black hole at $ T=0 $ to zero as dictated by the third law of thermodynamics. However, the main difference between condensed matter or cold atom systems and quantum black holes is the causality:  no horizon in the former separates the timelike from the spacelike.
The black hole has a horizon which is equal to its area,
so the third law of thermodynamics may need to be modified to consider this fact: 
its entropy for an extreme black hole at $ T=0 $  may still be non-zero because an observer outside the black hole can not see what is inside the black hole, the most he/she can see is the surface area of the black hole. 
Then one may need to modify the third law of thermodynamics to:  
the entropy of an extreme black hole at $ T=0 $ reaches its minimum, which is the surface of the black hole, namely, the BH entropy. If one squeezes the surface area to a point, then the BH entropy becomes zero,
and one recovers the third law of thermodynamics. 
It remains outstanding to evaluate the quantum corrections to the classical BH entropy for the extreme black holes, for example, in the context of 2d quantum gravity, which is dual to SYK models.
For a non-extreme black hole at any finite $ T $, then one may start to consider its Hawking radiations.

Note added: 
In a recent work \cite{QSLown}, we show that there is indeed a $ Z_2 $ QSL intervening
between the frustrated SF at weak coupling and the FM Mott at strong  coupling in Fig.\ref{condensate}(b).
So this work focus on symmetry broken states and the associated gapless excitations.
The approach used here is the microscopic calculations at both weak and strong couplings.
While \cite{QSLown} focus on topological states and the associated gapped fractionalized topological excitations.
The approach to be used there is a symmetry based effective action at any coupling strengths.
We will also establish the intrinsic relations between the two complementary approaches.

\acknowledgments
The authors thank 
Prof. Dapeng Yu for hospitality during the authors' visit at 
Institute for Quantum Science and Engineering, Shenzhen, China,
and thank Prof. Wei Ku for the hospitality during the authors visit at 
the Tsung-Dao Lee Institute, Shanghai, China.
This research was supported by AFOSR FA9550-16-1-0412.

\appendix

\section{The second step of the newly developed systematic
``order from quantum disorder analysis'' to compute the spectra
  of the slow-Goldstone mode: Canonical quantization approach }

In the Appendix A, we provide some technical details on
the second step of the newly developed systematic 
``order from quantum disorder analysis'' 
to compute the spectrum of the slow-Goldstone mode 
by the canonical quantization approach,
which is complementary to the boson coherent state path integral 
approach developed in the main text.
Reaching the same set of Goldstone modes by 
two independent and complementary approaches 
not only confirm the correctness of the results, 
but also provide additional physical insights into
the novel phenomenon of the slow-Goldstone mode generated by the OFQD.

In order to compute the corrections to the excitation spectrum due to the OFQD phenomena in the canonical quantization approach,
one need to express the order from disorder field operator $\delta\phi_\mathbf{r}$ in Eq.\eqref{B2}
in terms of the original boson field operator in Eq.\eqref{piflux}. 
The most generic state including all the possible quantum
fluctuations are parameterized in Eq.\eqref{state2}.
The $90^\circ$ quantum ground state corresponds to its saddle point value 
$ \Psi_0= (\rho_0,0,\pi/4,0,\chi_{1c},\chi_{2c}) $.
One can write down the most generic quantum fluctuations around the saddle point as
\begin{align}
    \Psi_i=\Psi_{i,0} + (\delta\rho,\theta,\delta\phi,\alpha,\delta\chi_{1},\delta\chi_{2})
\end{align}
The Bose field can be separated into the condensation part 
plus the quantum fluctuation part in a polar-like coordinate system:
\begin{align}
    \Psi_\mathbf{r}
	=\Psi_{\mathbf{r},0}
	+\delta\Psi_1 e^{-i\mathbf{K}\cdot\mathbf{r}}
	+\delta\Psi_2 e^{i\mathbf{K}\cdot\mathbf{r}}\>,
\end{align}
where the components at $ -\mathbf{K} $ and $ \mathbf{K} $ are:
\begin{eqnarray}
    \delta\Psi_1=
    \frac{\delta\rho-2\rho_0\delta\phi+i2\rho_0(\theta+\alpha)}
	 {2\sqrt{\rho_0}}
    (\eta_1\otimes\chi_{1c})          
    +   \sqrt{\rho_0}
    (\eta_1\otimes\delta\chi_1)        \nonumber   \\
    \delta\Psi_2=
    \frac{\delta\rho+2\rho_0\delta\phi+i2\rho_0(\theta-\alpha)}
	 {2\sqrt{\rho_0}}
    (\eta_2\otimes\chi_{2c})    
     +
    \sqrt{\rho_0}
    (\eta_2\otimes\delta\chi_2)
\end{eqnarray}

After comparing it with the decomposition in the Cartesian coordinate, using the fact
${\rm Re}[\chi_{1c}^\dagger\cdot\delta\chi_1]
={\rm Re}[\chi_{2c}^\dagger\cdot\delta\chi_2]=0 $
and introducing the $\xi_i$, $i =1,2 $ to be othor-normal to $\eta_i$, 
namely $ \xi^{\dagger}_i \eta_j= \delta_{ij} $,
One can express the order from disorder variable 
$\delta\phi_\mathbf{r}$ in terms of the original Bose fluctuation field:
\begin{align}
      \delta\phi_q
	=[&-\bar{\xi}_{1a}\delta a_{\mathbf{K+q}\uparrow}
	 -\xi_{1a}\delta a_{\mathbf{K-q}\uparrow}^\dagger
	 +\bar{\xi}_{2a}\delta a_{\mathbf{-K+q}\downarrow}   
	 +\xi_{2a}\delta a_{\mathbf{-K-q}\downarrow}^\dagger \nonumber   \\
	 &-\bar{\xi}_{1b}\delta b_{\mathbf{K+q}\uparrow}
	 -\xi_{1b}\delta b_{\mathbf{K-q}\uparrow}^\dagger    
	 +\bar{\xi}_{2b}\delta b_{\mathbf{-K+q}\downarrow}
	 +\xi_{2b}\delta b_{\mathbf{-K-q}\downarrow}^\dagger]/(4\sqrt{\rho_0})\>,
\end{align}
where as expected, both the quantum fluctuations near 
$ -\mathbf{K} $ and  $ \mathbf{K} $ in the Cartesian coordinate
appear in the right side of the equation.
Thus we can express the quantum corrections 
from the OFQD in terms of the original Bose fields as
\begin{align}
    \delta\mathcal{H}=\frac{B}{2} \sum_q\delta\phi_q\delta\phi_{-q}
\label{corr}
\end{align}

Finally, after combining Eq.\eqref{corr} with the $ \mathcal{H}^{(2)} $ in Eq.\eqref{h20},
we arrive at the effective Hamiltonian incorporating the order from quantum disorder (OFQD) mechanism:
\begin{align}
    H_\text{OFD}=H^{(2)}+\delta H=\frac{1}{2}\sum_{q}\Psi_q^\dagger (M+\delta M) \Psi_q
\label{total}
\end{align}
where the $ \delta M $  can be written as the following $ 8 \times 8 $ matrix.
\begin{align}
    \delta M=\frac{B}{16n_0}
	\begin{pmatrix}
	    1   &1  &e^{-i\pi/4} &e^{i\pi/4} &-1 &-1 &-e^{i\pi/4} &-e^{-i\pi/4}\\
	    1   &1	&e^{-i\pi/4} &e^{i\pi/4} &-1 &-1 &-e^{i\pi/4} &-e^{-i\pi/4}\\
	    e^{i\pi/4} &e^{i\pi/4} &1 &i &-e^{i\pi/4} &-e^{i\pi/4} &-i &-1\\
	    e^{-i\pi/4} &e^{-i\pi/4} &-i &1 &-e^{-i\pi/4} &-e^{-i\pi/4} &-1 &i\\
	    -1 &-1 &-e^{-i\pi/4} &-e^{i\pi/4} &1 &1 &e^{i\pi/4} &e^{-i\pi/4}\\
	    -1 &-1 &-e^{-i\pi/4} &-e^{i\pi/4} &1 &1 &e^{i\pi/4} &e^{-i\pi/4}\\
	    -e^{-i\pi/4} &-e^{-i\pi/4} &i &-1 &e^{-i\pi/4} &e^{-i\pi/4} &1 &-i\\
	    -e^{i\pi/4}  &-e^{i\pi/4}  &-1&-i &e^{i\pi/4}  &e^{i\pi/4}  &i &1\\
	\end{pmatrix}
\label{matrix88}
\end{align}

After applying Bogoliubov transformation to Eq.\eqref{matrix88}, 
we find the order from disorder correction Eq.\eqref{corr}
transfers the quadratic dispersion 
$ \Omega_2(\mathbf{q}) = \omega_{3}(\mathbf{q})= \frac{t}{\sqrt{2}}q^2  $ 
into a linear one
$ \omega_{3}=v_3 q=\sqrt{\frac{Bt}{2\sqrt{2}n_0}}q $.
   We also found that the SF Goldstone mode $ \Omega_1(\mathbf{q}) = \omega_{1}(\mathbf{q})= v_1 q  $  remains the same.
   Due to the momenta separation, $ \Omega_1( ( \pi,0)+ \mathbf{q} )= \omega_{2}(\mathbf{q})$ and
   $ \Omega_2( ( \pi,0)+ \mathbf{q} ) = \omega_{4}(\mathbf{q}) $ are not affected by the OFQD analysis.
   All the 4 linear modes were listed in Eq.\eqref{fourl}.

\section{The kinetic energies and currents in the 4-sublattice $ 90^{\circ} $ coplanar ground state.}

In the Appendix B, we evaluate the kinetic energy and current in the 4-sublattice $ 90^{\circ} $ coplanar quantum ground state.

 Using the method in \cite{yan-1,yan-2}, we will evaluate the conserved density-currents
 in the 4 sublattice $ 90^{\circ} $ coplanar ground state in Fig.\ref{classical}(a).
 In the context of 2d charge-vortex duality \cite{yan-1,yan-2},
 the vortex currents in the dual lattice gives the boson densities in the direct lattice.
 Here, we evaluate them directly on the direct lattice.

We can write the kinetic energy in the unit cell in Fig.\ref{fig:frog}(a) as:
\begin{align}
	&\mathcal{H}_\text{hop}
	=\sum_\mathbf{r}[ H_0+H_1+H_2+H_3+h.c.]
\end{align}
where $ H_0=t a_{\mathbf{r}\sigma}^\dagger b_{\mathbf{r}\sigma},
	H_1=-t a_{\mathbf{r}\sigma}^\dagger b_{\mathbf{r+v}_1,\sigma},
	H_2=-t a_{\mathbf{r}\sigma}^\dagger b_{\mathbf{r+v}_2,\sigma},
	H_3=-t a_{\mathbf{r}\sigma}^\dagger b_{\mathbf{r}+\mathbf{v}_1+\mathbf{v}_2,\sigma} $
and the two lattice vectors in Fig.\ref{fig:frog}(a) are $ \mathbf{v}_1=(1,1),~~\mathbf{v}_2=(-1,1) $.
Obviously, in the $ Z_2 $ gauge, the first bond is frustrated, the other three are not.
Along a given bond $ (i, i+ \mu) $:
\begin{equation}
     H_\mu =K_\mu-i I_\mu\>,
\end{equation}
where $ K $ is the kinetic energy 
and $ I $ is the current flowing along the bond \cite{yan-1,yan-2}.

The mean-field wave-function of the 4-sublattice $ 90^{\circ} $ coplanar ground state  is given in Eq.\eqref{state0} with $ \phi=\pi/4 $
 and $\mathbf{r}$ labels the unit cell which contains both $ A $ and $ B $ sublattice sites.
 Since any nearest neighbor inside a unit cell consists of A-B sublattice sites,
 we only need to evaluate the form $ a_\mathbf{r}^\dagger b_{\mathbf{r}+\boldsymbol{\mu}} $.
 Substituting the wave-function into the form leads to:
\begin{align}
    a_\mathbf{r}^\dagger b_{\mathbf{r}+\boldsymbol{\mu}}
	=-{\rm Re}[n_0e^{i\frac{\pi}{4}}e^{i\mathbf{K}\cdot\boldsymbol{\mu}}]\>,
\end{align}
 which simplifies to
\begin{align}
    H_0=H_1=H_2=H_3=-n_0t/\sqrt{2}
\end{align}
  It shows that the currents vanish and
  the kinetic energies are uniform in the ground state.
  If the system has open edges, then there is no edge states.
The system just arranges itself into the 4-sublattice $ 90^{\circ} $ coplanar ground state to offset
  the external $ \pi $ flux and lead to no currents and  no frustrated kinetic energy bonds.
  If the system has edges, then there is no edge currents flowing along the edges.

\bibliographystyle{JHEP}
\bibliography{Goldstone.bib}

\providecommand{\href}[2]{#2}\begingroup\raggedright\begin{thebibliography}{10}

\bibitem{GS1}
Y.~Nambu, \emph{Quasi-particles and gauge invariance in the theory of
  superconductivity},
  \href{https://doi.org/10.1103/PhysRev.117.648}{\emph{Phys. Rev.} {\bfseries
  117} (1960) 648}.

\bibitem{GS2}
J.~Goldstone, \emph{{Field Theories with Superconductor Solutions}},
  \href{https://doi.org/10.1007/BF02812722}{\emph{Nuovo Cim.} {\bfseries 19}
  (1961) 154}.

\bibitem{GS3}
J.~Goldstone, A.~Salam and S.~Weinberg, \emph{Broken symmetries},
  \href{https://doi.org/10.1103/PhysRev.127.965}{\emph{Phys. Rev.} {\bfseries
  127} (1962) 965}.

\bibitem{Stringbook}
K.~Becker, M.~Becker and J.H.~Schwarz, \emph{String Theory and M-Theory: A
  Modern Introduction}, Cambridge University Press (2006),
  \href{https://doi.org/10.1017/CBO9780511816086}{10.1017/CBO9780511816086}.

\bibitem{Stringbook2}
E.~Kiritsis, \emph{String Theory in a Nutshell}, Cambridge University Press,
  1st~ed. (2007).

\bibitem{SY}
S.~Sachdev and J.~Ye, \emph{Gapless spin-fluid ground state in a random quantum
  heisenberg magnet},
  \href{https://doi.org/10.1103/PhysRevLett.70.3339}{\emph{Phys. Rev. Lett.}
  {\bfseries 70} (1993) 3339}.

\bibitem{Kit}
A.~Kitaev, ``A simple model of quantum holography.'' April 7, 2015 and May 27,
  2015.

\bibitem{Mald}
J.~Maldacena and D.~Stanford, \emph{Remarks on the {Sachdev-Ye-Kitaev} model},
  \href{https://doi.org/10.1103/PhysRevD.94.106002}{\emph{Phys. Rev. D}
  {\bfseries 94} (2016) 106002}.

\bibitem{QFT}
M.E.~Peskin and D.V.~Schroeder, \emph{{An introduction to quantum field
  theory}}, Westview, Boulder, CO (1995).

\bibitem{qaf1}
S.~Sachdev and J.~Ye, \emph{Universal quantum-critical dynamics of
  two-dimensional antiferromagnets},
  \href{https://doi.org/10.1103/PhysRevLett.69.2411}{\emph{Phys. Rev. Lett.}
  {\bfseries 69} (1992) 2411}.

\bibitem{qaf2}
A.V.~Chubukov, S.~Sachdev and J.~Ye, \emph{Theory of two-dimensional quantum
  heisenberg antiferromagnets with a nearly critical ground state},
  \href{https://doi.org/10.1103/PhysRevB.49.11919}{\emph{Phys. Rev. B}
  {\bfseries 49} (1994) 11919}.

\bibitem{u1largen}
J.~Ye and C.~Zhang, \emph{{Superradiance, Berry phase, photon phase diffusion,
  and number squeezed state in the {U(1)} Dicke (Tavis-Cummings) model}},
  \href{https://doi.org/10.1103/PhysRevA.84.023840}{\emph{Phys. Rev. A}
  {\bfseries 84} (2011) 023840}.

\bibitem{gold}
Y.-X.~Yu, J.~Ye and W.-M.~Liu, \emph{{Goldstone and Higgs modes of photons
  inside a cavity}}, \href{https://doi.org/10.1038/srep03476}{\emph{Scientific
  Reports} {\bfseries 3} (2013) 3476}.

\bibitem{strongED}
Y.~Yi-Xiang, J.~Ye and C.~Zhang, \emph{Parity oscillations and photon
  correlation functions in the ${Z}_{2}\text{\ensuremath{-}}\mathrm{U}(1)$
  {Dicke} model at a finite number of atoms or qubits},
  \href{https://doi.org/10.1103/PhysRevA.94.023830}{\emph{Phys. Rev. A}
  {\bfseries 94} (2016) 023830}.

\bibitem{QIbook}
B.~Zeng, X.~Chen, D.-L.~Zhou and X.-G.~Wen, \emph{Quantum Information Meets
  Quantum Matter}, Springer-Verlag New York (2019),
  \href{https://doi.org/10.1007/978-1-4939-9084-9}{10.1007/978-1-4939-9084-9}.

\bibitem{CWpotential}
S.~Coleman and E.~Weinberg, \emph{Radiative corrections as the origin of
  spontaneous symmetry breaking},
  \href{https://doi.org/10.1103/PhysRevD.7.1888}{\emph{Phys. Rev. D} {\bfseries
  7} (1973) 1888}.

\bibitem{CW2}
S.~Weinberg, \emph{Approximate symmetries and pseudo-goldstone bosons},
  \href{https://doi.org/10.1103/PhysRevLett.29.1698}{\emph{Phys. Rev. Lett.}
  {\bfseries 29} (1972) 1698}.

\bibitem{weinberg_1996}
S.~Weinberg, \emph{The Quantum Theory of Fields}, vol.~2, Cambridge University
  Press (1996),
  \href{https://doi.org/10.1017/CBO9781139644174}{10.1017/CBO9781139644174}.

\bibitem{subir}
S.~Sachdev, \emph{Kagome- and triangular-lattice heisenberg antiferromagnets:
  Ordering from quantum fluctuations and quantum-disordered ground states with
  unconfined bosonic spinons},
  \href{https://doi.org/10.1103/PhysRevB.45.12377}{\emph{Phys. Rev. B}
  {\bfseries 45} (1992) 12377}.

\bibitem{gan}
G.~Murthy, D.~Arovas and A.~Auerbach, \emph{Superfluids and supersolids on
  frustrated two-dimensional lattices},
  \href{https://doi.org/10.1103/PhysRevB.55.3104}{\emph{Phys. Rev. B}
  {\bfseries 55} (1997) 3104}.

\bibitem{frusrev}
C.~Lhuillier and G.~Misguich, \emph{Frustrated quantum magnets},
  2001,arXiv:cond-mat/0109146.

\bibitem{sachdev}
S.~Sachdev, \emph{Quantum Phase Transitions}, Cambridge University Press,
  2nd~ed. (2011),
  \href{https://doi.org/10.1017/CBO9780511973765}{10.1017/CBO9780511973765}.

\bibitem{Sachdev1991}
S.~Sachdev and N.~Read, \emph{Large n expansion for frustrated and doped
  quantum antiferromagnets},
  \href{https://doi.org/10.1142/S0217979291000158}{\emph{International Journal
  of Modern Physics B} {\bfseries 05} (1991) 219}.

\bibitem{Balents2010}
L.~Balents, \emph{Spin liquids in frustrated magnets},
  \href{https://doi.org/10.1038/nature08917}{\emph{Nature} {\bfseries 464}
  (2010) 199}.

\bibitem{FrustratedBook}
C.~Lacroix, P.~Mendels and F.~Mila, \emph{Introduction to Frustrated Magnetism:
  Materials, Experiments, Theory}, Springer-Verlag Berlin Heidelberg (2011),
  \href{https://doi.org/10.1007/978-3-642-10589-0}{10.1007/978-3-642-10589-0}.

\bibitem{stagg1}
M.~Aidelsburger, M.~Atala, S.~Nascimb\`ene, S.~Trotzky, Y.-A.~Chen and
  I.~Bloch, \emph{Experimental realization of strong effective magnetic fields
  in an optical lattice},
  \href{https://doi.org/10.1103/PhysRevLett.107.255301}{\emph{Phys. Rev. Lett.}
  {\bfseries 107} (2011) 255301}.

\bibitem{stagg2-1}
J.~Struck, C.~{\"O}lschl{\"a}ger, R.~Le~Targat, P.~Soltan-Panahi, A.~Eckardt,
  M.~Lewenstein et~al., \emph{Quantum simulation of frustrated classical
  magnetism in triangular optical lattices},
  \href{https://doi.org/10.1126/science.1207239}{\emph{Science} {\bfseries 333}
  (2011) 996}
  [\href{https://arxiv.org/abs/https://science.sciencemag.org/content/333/6045/996.full.pdf}{{\ttfamily
  https://science.sciencemag.org/content/333/6045/996.full.pdf}}].

\bibitem{stagg2-2}
J.~Struck, C.~\"Olschl\"ager, M.~Weinberg, P.~Hauke, J.~Simonet, A.~Eckardt
  et~al., \emph{Tunable gauge potential for neutral and spinless particles in
  driven optical lattices},
  \href{https://doi.org/10.1103/PhysRevLett.108.225304}{\emph{Phys. Rev. Lett.}
  {\bfseries 108} (2012) 225304}.

\bibitem{stagg2-3}
J.~Struck, M.~Weinberg, C.~\"Olschl\"ager, P.~Windpassinger, J.~Simonet,
  K.~Sengstock et~al., \emph{{Engineering Ising-XY spin-models in a triangular
  lattice using tunable artificial gauge fields}},
  \href{https://doi.org/10.1038/nphys2750}{\emph{Nature Physics} {\bfseries 9}
  (2012) 738}.

\bibitem{uniform1}
M.~Aidelsburger, M.~Atala, M.~Lohse, J.T.~Barreiro, B.~Paredes and I.~Bloch,
  \emph{{Realization of the Hofstadter Hamiltonian with Ultracold Atoms in
  Optical Lattices}},
  \href{https://doi.org/10.1103/PhysRevLett.111.185301}{\emph{Phys. Rev. Lett.}
  {\bfseries 111} (2013) 185301}.

\bibitem{uniform2}
H.~Miyake, G.A.~Siviloglou, C.J.~Kennedy, W.C.~Burton and W.~Ketterle,
  \emph{Realizing the harper hamiltonian with laser-assisted tunneling in
  optical lattices},
  \href{https://doi.org/10.1103/PhysRevLett.111.185302}{\emph{Phys. Rev. Lett.}
  {\bfseries 111} (2013) 185302}.

\bibitem{uniform3}
C.J.~Kennedy, G.A.~Siviloglou, H.~Miyake, W.C.~Burton and W.~Ketterle,
  \emph{Spin-orbit coupling and quantum spin hall effect for neutral atoms
  without spin flips},
  \href{https://doi.org/10.1103/PhysRevLett.111.225301}{\emph{Phys. Rev. Lett.}
  {\bfseries 111} (2013) 225301}.

\bibitem{newexp}
M.~Atala, M.~Aidelsburger, M.~Lohse, J.T.~Barreiro1, B.~Paredes and I.~Bloch,
  \emph{Observation of chiral currents with ultracold atoms in bosonic
  ladders}, \href{https://doi.org/10.1038/nphys2998}{\emph{Nature Physics}
  {\bfseries 10} (2014) 588}.

\bibitem{newexp2}
M.~Aidelsburger, M.~Lohse, C.~Schweizer, M.~Atala, J.T.~Barreiro, S.~Nascimbe
  et~al., \emph{{Measuring the Chern number of Hofstadter bands with ultracold
  bosonic atoms}}, .

\bibitem{haldane}
G.~Jotzu, M.~Messer, R.~Desbuquois, M.~Lebrat, T.~Uehlinger, D.~Greif et~al.,
  \emph{{Experimental realization of the topological Haldane model with
  ultracold fermions}},
  \href{https://doi.org/10.1038/nature13915}{\emph{Nature} {\bfseries 515}
  (2014) 237}.

\bibitem{quater1}
C.~Owens, A.~LaChapelle, B.~Saxberg, B.M.~Anderson, R.~Ma, J.~Simon et~al.,
  \emph{Quarter-flux {Hofstadter} lattice in a qubit-compatible microwave
  cavity array}, \href{https://doi.org/10.1103/PhysRevA.97.013818}{\emph{Phys.
  Rev. A} {\bfseries 97} (2018) 013818}.

\bibitem{quater2}
B.M.~Anderson, R.~Ma, C.~Owens, D.I.~Schuster and J.~Simon, \emph{Engineering
  topological many-body materials in microwave cavity arrays},
  \href{https://doi.org/10.1103/PhysRevX.6.041043}{\emph{Phys. Rev. X}
  {\bfseries 6} (2016) 041043}.

\bibitem{quater3}
J.~Ningyuan, C.~Owens, A.~Sommer, D.~Schuster and J.~Simon, \emph{Time- and
  site-resolved dynamics in a topological circuit},
  \href{https://doi.org/10.1103/PhysRevX.5.021031}{\emph{Phys. Rev. X}
  {\bfseries 5} (2015) 021031}.

\bibitem{quater4}
N.~Schine, A.~Ryou, A.~Gromov, A.~Sommer and J.~Simon, \emph{{Synthetic Landau
  levels for photons}},
  \href{https://doi.org/10.1038/nature17943}{\emph{Nature} {\bfseries 534}
  (2016) 671}.

\bibitem{QSLown}
F.~Sun and J.~Ye, \emph{Quantum spin liquids in a square lattice subject to an
  abelian flux and its experimental observation},  2019, arXiv:1903.11134.

\bibitem{rh}
F.~Sun, J.~Ye and W.-M.~Liu, \emph{Quantum magnetism of spinor bosons in
  optical lattices with synthetic non-abelian gauge fields},
  \href{https://doi.org/10.1103/PhysRevA.92.043609}{\emph{Phys. Rev. A}
  {\bfseries 92} (2015) 043609}.

\bibitem{hh-1}
D.R.~Hofstadter, \emph{Energy levels and wave functions of bloch electrons in
  rational and irrational magnetic fields},
  \href{https://doi.org/10.1103/PhysRevB.14.2239}{\emph{Phys. Rev. B}
  {\bfseries 14} (1976) 2239}.

\bibitem{hh-2}
J.~Zak, \emph{Magnetic translation group},
  \href{https://doi.org/10.1103/PhysRev.134.A1602}{\emph{Phys. Rev.} {\bfseries
  134} (1964) A1602}.

\bibitem{hh-3}
J.~Zak, \emph{Magnetic translation group. ii. irreducible representations},
  \href{https://doi.org/10.1103/PhysRev.134.A1607}{\emph{Phys. Rev.} {\bfseries
  134} (1964) A1607}.

\bibitem{BLgraphene}
C.R.~Dean, L.~Wang, P.~Maher, C.~Forsythe, F.~Ghahari, Y.~Gao et~al.,
  \emph{Hofstadter’s butterfly and the fractal quantum hall effect in moiré
  superlattices}, \href{https://doi.org/10.1038/nature12186}{\emph{Nature}
  {\bfseries 497} (2013) 598–602}.

\bibitem{blochrmp}
I.~Bloch, J.~Dalibard and W.~Zwerger, \emph{Many-body physics with ultracold
  gases}, \href{https://doi.org/10.1103/RevModPhys.80.885}{\emph{Rev. Mod.
  Phys.} {\bfseries 80} (2008) 885}.

\bibitem{braggbog-1}
M.~Kozuma, L.~Deng, E.W.~Hagley, J.~Wen, R.~Lutwak, K.~Helmerson et~al.,
  \emph{Coherent splitting of bose-einstein condensed atoms with optically
  induced bragg diffraction},
  \href{https://doi.org/10.1103/PhysRevLett.82.871}{\emph{Phys. Rev. Lett.}
  {\bfseries 82} (1999) 871}.

\bibitem{braggbog-2}
J.~Stenger, S.~Inouye, A.P.~Chikkatur, D.M.~Stamper-Kurn, D.E.~Pritchard and
  W.~Ketterle, \emph{Bragg spectroscopy of a bose-einstein condensate},
  \href{https://doi.org/10.1103/PhysRevLett.82.4569}{\emph{Phys. Rev. Lett.}
  {\bfseries 82} (1999) 4569}.

\bibitem{braggbog-3}
D.M.~Stamper-Kurn, A.P.~Chikkatur, A.~G\"orlitz, S.~Inouye, S.~Gupta,
  D.E.~Pritchard et~al., \emph{Excitation of phonons in a bose-einstein
  condensate by light scattering},
  \href{https://doi.org/10.1103/PhysRevLett.83.2876}{\emph{Phys. Rev. Lett.}
  {\bfseries 83} (1999) 2876}.

\bibitem{braggbog-4}
J.~Steinhauer, R.~Ozeri, N.~Katz and N.~Davidson, \emph{Excitation spectrum of
  a bose-einstein condensate},
  \href{https://doi.org/10.1103/PhysRevLett.88.120407}{\emph{Phys. Rev. Lett.}
  {\bfseries 88} (2002) 120407}.

\bibitem{braggbog-5}
S.B.~Papp, J.M.~Pino, R.J.~Wild, S.~Ronen, C.E.~Wieman, D.S.~Jin et~al.,
  \emph{Bragg spectroscopy of a strongly interacting $^{85}\mathrm{Rb}$
  bose-einstein condensate},
  \href{https://doi.org/10.1103/PhysRevLett.101.135301}{\emph{Phys. Rev. Lett.}
  {\bfseries 101} (2008) 135301}.

\bibitem{braggangle}
P.T.~Ernst, S.~G\:otze, J.S.~Krauser, K.~Pyka, D.-S.~L\"uhmann, D.~Pfannkuche
  et~al., \emph{Probing superfluids in optical lattices by momentum-resolved
  bragg spectroscopy}, \href{https://doi.org/10.1038/nphys1476}{\emph{Nature
  Physics} {\bfseries 6} (2009) 56–61}.

\bibitem{braggeng}
T.~St\"oferle, H.~Moritz, C.~Schori, M.~K\"ohl and T.~Esslinger,
  \emph{Transition from a strongly interacting 1d superfluid to a mott
  insulator}, \href{https://doi.org/10.1103/PhysRevLett.92.130403}{\emph{Phys.
  Rev. Lett.} {\bfseries 92} (2004) 130403}.

\bibitem{braggsingle-1}
G.~Birkl, M.~Gatzke, I.H.~Deutsch, S.L.~Rolston and W.D.~Phillips, \emph{Bragg
  scattering from atoms in optical lattices},
  \href{https://doi.org/10.1103/PhysRevLett.75.2823}{\emph{Phys. Rev. Lett.}
  {\bfseries 75} (1995) 2823}.

\bibitem{braggsingle-2}
M.~Weidem\"uller, A.~Hemmerich, A.~G\"orlitz, T.~Esslinger and T.W.~H\"ansch,
  \emph{Bragg diffraction in an atomic lattice bound by light},
  \href{https://doi.org/10.1103/PhysRevLett.75.4583}{\emph{Phys. Rev. Lett.}
  {\bfseries 75} (1995) 4583}.

\bibitem{braggsingle-3}
J.~Ruostekoski, C.J.~Foot and A.B.~Deb, \emph{Light scattering for thermometry
  of fermionic atoms in an optical lattice},
  \href{https://doi.org/10.1103/PhysRevLett.103.170404}{\emph{Phys. Rev. Lett.}
  {\bfseries 103} (2009) 170404}.

\bibitem{becbragg}
S.-C.~Ji, L.~Zhang, X.-T.~Xu, Z.~Wu, Y.~Deng, S.~Chen et~al., \emph{Softening
  of roton and phonon modes in a bose-einstein condensate with spin-orbit
  coupling}, \href{https://doi.org/10.1103/PhysRevLett.114.105301}{\emph{Phys.
  Rev. Lett.} {\bfseries 114} (2015) 105301}.

\bibitem{bragg12-1}
J.~Ye, J.M.~Zhang, W.M.~Liu, K.~Zhang, Y.~Li and W.~Zhang,
  \emph{Light-scattering detection of quantum phases of ultracold atoms in
  optical lattices},
  \href{https://doi.org/10.1103/PhysRevA.83.051604}{\emph{Phys. Rev. A}
  {\bfseries 83} (2011) 051604}.

\bibitem{bragg12-2}
J.~Ye, K.~Zhang, Y.~Li, Y.~Chen and W.~Zhang, \emph{Optical bragg, atomic bragg
  and cavity qed detections of quantum phases and excitation spectra of
  ultracold atoms in bipartite and frustrated optical lattices},
  \href{https://doi.org/10.1016/j.aop.2012.09.006}{\emph{Annals of Physics}
  {\bfseries 328} (2013) 103–138}.

\bibitem{SS1}
M.~Endres, T.~Fukuhara, D.~Pekker, M.~Cheneau, P.~Schau$\beta$, C.~Gross
  et~al., \emph{{The `Higgs' amplitude mode at the two-dimensional
  superfluid/Mott insulator transition}},
  \href{https://doi.org/10.1038/nature11255}{\emph{Nature} {\bfseries 487}
  (2012) 454–}.

\bibitem{SS2}
J.~L{\'e}onard, A.~Morales, P.~Zupancic, T.~Donner and T.~Esslinger,
  \emph{{Monitoring and manipulating Higgs and Goldstone modes in a supersolid
  quantum gas}}, \href{https://doi.org/10.1126/science.aan2608}{\emph{Science}
  {\bfseries 358} (2017) 1415}.

\bibitem{axion}
R.D.~Peccei, \emph{The strong $\mathrm{CP}$ problem and axions},
  \href{https://doi.org/10.1007/978-3-540-73518-2_1}{\emph{Axions} (2008)
  3–17}.

\bibitem{axion2}
R.D.~Peccei and H.R.~Quinn, \emph{$\mathrm{CP}$ conservation in the presence of
  pseudoparticles},
  \href{https://doi.org/10.1103/PhysRevLett.38.1440}{\emph{Phys. Rev. Lett.}
  {\bfseries 38} (1977) 1440}.

\bibitem{axion3}
R.D.~Peccei and H.R.~Quinn, \emph{Constraints imposed by $\mathrm{CP}$
  conservation in the presence of pseudoparticles},
  \href{https://doi.org/10.1103/PhysRevD.16.1791}{\emph{Phys. Rev. D}
  {\bfseries 16} (1977) 1791}.

\bibitem{yan-1}
Y.~Chen and J.~Ye, \emph{Characterizing symmetry breaking patterns in a lattice
  by dual degrees of freedom},
  \href{https://doi.org/10.1080/14786435.2012.712221}{\emph{Philosophical
  Magazine} {\bfseries 92} (2012) 4484–4491}.

\bibitem{yan-2}
J.~Ye and Y.~Chen, \emph{Quantum phases, supersolids and quantum phase
  transitions of interacting bosons in frustrated lattices},
  \href{https://doi.org/10.1016/j.nuclphysb.2012.11.022}{\emph{Nuclear Physics
  B} {\bfseries 869} (2013) 242–281}.

\end{thebibliography}\endgroup

\end{document}